\documentclass[a4paper,12pt]{article}

\newcommand{\sect}[1]{\setcounter{equation}{0}\section{#1}}

\begin{document}
\topmargin 0pt
\oddsidemargin 0mm

\renewcommand{\thefootnote}{\fnsymbol{footnote}}
\begin{titlepage}
\begin{flushright}
hep-th/0112253
\end{flushright}

\vspace{5mm}
\begin{center}
{\Large \bf Cardy-Verlinde Formula and Thermodynamics of Black Holes in de
 Sitter Spaces}
\vspace{12mm}

{\large
Rong-Gen Cai\footnote{Email address: cairg@itp.ac.cn}\\
\vspace{8mm}
{ \em Institute of Theoretical Physics, Chinese Academy of Sciences, \\
   P.O. Box 2735, Beijing 100080, China}} 
\end{center}
\vspace{5mm}
\centerline{{\bf{Abstract}}}
\vspace{5mm}
We continue the study of thermodynamics of black holes in de Sitter spaces. In a previous
paper (hep-th/0111093), we have shown that the entropy of cosmological horizon in the 
Schwarzschild-de Sitter solutions and topological de Sitter solutions can be expressed in
a form of the Cardy-Verlinde formula, if one adopts the prescription to compute the 
gravitational mass from data at early or late time infinity of de Sitter space. However,
this definition of gravitational mass cannot give a similar expression like the 
Cardy-Verlinde formula for the entropy associated with the horizon of black holes in de 
Sitter spaces. In this paper, we first generalize the previous discussion to the cases of  
Reissner-Nordstr\"om-de Sitter solutions and Kerr-de Sitter solutions. Furthermore, we find that
the entropy of black hole horizon can also be rewritten in terms of the Cardy-Verlinde
formula for these black holes in de Sitter spaces, if we use the definition due to Abbott and
Deser for conserved charges in asymptotically de Sitter spaces. We discuss the 
implication of our result. In addition, we give the first law of de Sitter black hole 
mechanics.

\end{titlepage}

\newpage
\renewcommand{\thefootnote}{\arabic{footnote}}
\setcounter{footnote}{0}
\setcounter{page}{2}

\sect{Introduction}
Recently much attention has been focused on studying the de Sitter (dS) space and 
asymptotically dS space. This is motivated at least by the following two aspects.
First, recent analysis of astronomical data for supernova indicates that there is a
positive cosmological constant in our universe \cite{Per,CDS,Gar}. Thus our universe
might approach to a dS phase in the far future \cite{HKS,FKMP}. Second, defined in 
a manner analogous to the AdS/CFT correspondence, an interesting proposal, the so-called
dS/CFT correspondence, has been 
suggested recently that there is a dual between quantum gravity on a dS space and 
a Euclidean conformal field (CFT) on a boundary of the dS space \cite{Stron1}
(for earlier works on this proposal see \cite{Hull,Bala,Witten1,Mazu}).

Unlike the case of asymptotically flat and asymptotically AdS spacetimes, however,
it is not an easy matter to compute conserved charges associated with an asymptotically 
dS space because of the absence of spatial infinity and a globally timelike Killing 
vector in such a spacetime. Up to the best of the present author's knowledge, there are
two prescriptions to calculate conserved charges of asymptotically dS spaces.  One of them
is the prescription proposed  recently by Balasubranmanian, de Boer 
and Minic (BBM) \cite{BBM}. Using this prescription together with the surface counterterm 
method\footnote{The surface
counterterm method in the asymptotically dS space has been also discussed 
in~\cite{Noji1, Klem1}.}, one can compute the boundary stress-energy tensor and
conserved charges of asymptotically dS spaces from data at early or late time infinity.   
In this way the authors of \cite{BBM} calculated the masses of three-, four- and 
five-dimensional Schwarzschild-dS (SdS) solutions. It is found that pure dS spaces are 
always more massive than SdS solutions in the corresponding dimensions: subtracting the
anomalous Casimir energy of pure dS spaces in odd dimensions (in even dimensional dS spaces
there is no associated  anomalous Casimir energy), the masses of SdS black hole 
solutions are always negative. This is also confirmed in higher dimensional SdS
solutions~\cite{Mann} and Kerr-dS solutions~\cite{Dehg}.  On the basis of this result,
 the authors of \cite{BBM} proposed an intriguing conjecture:{\it Any asymptotically dS space 
whose mass exceeds that of pure dS space contains a cosmological singularity.} This conjecture is
 partially verified within the topological dS solution and its dilatonic 
deformation in~\cite{CMZ}.

Adopting the definition of mass in the BBM prescription, in a previous paper~\cite{Cai1},
we have shown that the entropy of cosmological horizons in the SdS  solutions
and topological dS solutions can be expressed in terms of the Cardy-Verlinde 
formula~\cite{Verl}, which is supposed to be an entropy formula of CFT in any 
dimension\footnote{In the AdS/CFT correspondence, the Cardy-Verlinde formula has
been shown to hold in many cases, for instance, Schwarzschild-AdS black holes~\cite{Verl},
Kerr-AdS black holes~\cite{Klem2}, hyperbolic AdS black holes~\cite{Cai2}, charged AdS
black holes~\cite{Cai2} and the Taub-Bolt AdS instanton solutions~\cite{Birm}. In the
cosmological context, there are a lot of works on the discussion of the Cardy-Verlinde
formula, for a complete list, for example, see the more recent reference \cite{Myung1}.}.
Thus our result provides support of the dS/CFT correspondence. However, there are two
points to be well understood. The first is that the mass (energy of the dual CFT) is 
negative, and it is measured at the far past (${\cal I}^-$) or far future (${\cal I}^+$)
 boundary of dS space, which is outside the cosmological horizon. Therefore, the boundary is
not accessible for observers inside the cosmological horizon. The other is that for 
the SdS black hole spacetime, except for the cosmological horizon,
there is a black hole horizon, which has also associated  Hawking radiation and 
entropy; the spacetime is believed to have total entropy which is the sum of black hole horizon
entropy and cosmological horizon entropy\footnote{Even in the thermodynamic sense,
there is no well-defined derivation for the total entropy being the sum of the black hole
horizon entropy and cosmological horizon entropy. But when the temperature of black  hole 
horizon is equal to that of cosmological horizon, it can be shown,
for example see \cite{CJS} and references therein.}. In Ref.~\cite{Cai1}, it is found that
if one  adopts the mass definition in the BBM prescription, the entropy of black hole horizon
cannot be rewritten in a form like the Cardy-Verlinde formula.

The other prescription to compute conserved charges of asymptotically dS spaces is developed
by Abbott and Deser (AD) \cite{AD}, by considering the deviation of metric from the pure
dS space being defined as the vacuum (lowest energy state). In the AD prescription,
 the gravitational
mass of asymptotically dS spaces is alway positive, and coincides with the ADM mass in 
asymptotically flat spacetimes, when the cosmological constant goes to zero.  

In this paper, we will generalize the discussion in \cite{Cai1} to the cases of 
Reissner-Norstr\"om-dS (RNdS) and Kerr-dS (KdS) spacetimes. That is, we will show that the 
entropy of cosmological horizon in the RNdS and KdS solutions can also be rewritten in the form
of Cardy-Verlinde formula.  We then show that if one uses the AD prescription, the entropy of 
black hole horizons in dS spaces can also be expressed by the Cardy-Verlinde formula, but the 
extensive part of energy is found to be negative. We will also discuss the first law of dS 
black hole solutions. The organization of the paper is as follows. In Sec.~2, 3, 4 we will 
consider the SdS, RNdS, and KdS black hole spacetimes, respectively. The case of pure dS space
in a rotating coordinate system  will also be discussed in Sec.~4.
In Sec.~5 we summary  and discuss our results.

\sect{Schwarzschild-de Sitter Black Holes }

We start with an $(n+2)$-dimensional SdS black hole solution, whose metric is
\begin{equation}
\label{2eq1}
ds^2 =-f(r)dt^2 +f(r)^{-1}dr^2 +r^2 d\Omega_n^2,
\end{equation}
where
\begin{equation}
f(r) = 1 -\frac{\omega_n M}{r^{n-1}} -\frac{r^2}{l^2}, 
\ \ \omega_n=\frac{16\pi G}{n\mbox {Vol}(S^n)}. 
\end{equation}
Here $G$ is the gravitational constant in $(n+2)$ dimensions, $l$ is the curvature radius
of dS space, $\mbox{Vol}(S^n)$ denotes the volume of a unit $n$-sphere $d\Omega_n^2$, and
$M$ is an integration constant. When $M=0$, the solution (\ref{2eq1}) reduces to the pure
dS space with a cosmological horizon at $r_c=l$. When $M$ increases with $M>0$, a black hole
horizon occurs and increases in size with $M$, while the cosmological horizon shrinks.
Finally the black hole horizon $r_+$ touches the cosmological horizon $r_c$ when
\begin{equation}
\label{2eq3}
M= M_N \equiv \frac{2}{\omega_n (n+1)}\left(\frac{n-1}{n+1}l^2\right)^{(n-1)/2}.
\end{equation}
This is the Nariai black hole, the maximal black hole in dS space. When $M>M_N$, both the
two horizons disappear and the solution (\ref{2eq1}) describes a naked singularity.  
The cosmological horizon $r_c$ and black hole horizon $r_+$ are two positive 
real roots of the 
equation, $f(r)=0$. The cosmological horizon is the larger one  and the black hole horizon
is smaller one. The thermodynamics associated with the cosmological horizon has been discussed
in \cite{Cai1}. For completeness and convenience for discussing the RNdS and KdS cases below,
we briefly give main properties here. The cosmological horizon of the SdS solution has
the Hawking temperature $T$ and entropy $S$,
\begin{eqnarray}
\label{2eq4}
&&  T=\frac{1}{4\pi r_c}\left((n+1)\frac{r_c^2}{l^2} -(n-1)\right), \nonumber \\
&& S=\frac{r_c^n \mbox{Vol}(S^n)}{4G}.
\end{eqnarray}
Subtracting the anomalous Casimir energy of pure dS spaces in odd dimensions, in the BBM 
prescription the gravitational mass of the SdS black holes is~\cite{BBM,Mann} 
\begin{equation}
\label{2eq5}
E=-M=\frac{r_c^{n-1}}{\omega_n}\left(\frac{r_c^2}{l^2}-1\right), 
\end{equation}
where the mass is expressed in terms of the cosmological horizon radius $r_c$. Following
Ref.~\cite{Verl}, the Casimir energy $E_c$ (non-extensive part of total energy), defined 
as $ E_c=(n+1)E-nTS$, is found to be~\cite{Cai1}
\begin{equation}
\label{2eq6}
E_c=-\frac{2nr_c^{n-1}\mbox{Vol}(S^n)}{16\pi G}.
\end{equation}
The negative definite Casimir energy is consistent with the argument that in the dS/CFT 
correspondence, the dual CFT is not unitary~\cite{Stron1}. Further, one has 
\begin{equation}
2E-E_c=\frac{2n r_c^{n+1}\mbox{Vol}(S^n)}{16\pi G l^2}, 
\end{equation}
and  the entropy $S$ (\ref{2eq4}) of the cosmological horizon can be rewritten 
as \cite{Cai1}\footnote{For related discussions, see also \cite{Dan,Halyo}, from another angle.}
\begin{equation}
\label{2eq8}
S=\frac{2\pi l}{n}\sqrt{|E_c|(2E-E_c)}, 
\end{equation}
a form of the Cardy-Verlinde formula~\cite{Verl}. This result (\ref{2eq8}) provides 
evidence that the thermodynamics of cosmological horizon in the SdS solution can be 
described by a CFT. In other words, our result is in favor of the dS/CFT 
correspondence. For pure dS space, one has $E=0$ and $r_c=l$, the formula (\ref{2eq8})
precisely reproduces the entropy of pure dS space as well. 

In addition, it is easy to check that the BBM mass (\ref{2eq5}), Hawking temperature $T$
and entropy $S$ in (\ref{2eq4}) of the cosmological horizon satisfy the first law of 
thermodynamics 
\begin{equation}
\label{2eq9}
dE=TdS.
\end{equation}
On the other hand, the black hole horizon $r_+$ in the SdS solution has also associated 
Hawking temperature $\tilde T$ and entropy $\tilde S$~\footnote{Throughout this paper,
notations with tildes will denote quantities associated with the black hole horizon.}
\begin{eqnarray}
\label{2eq10}
&& \tilde T=\frac{1}{4\pi r_+}\left ( (n-1) -(n+1)\frac{r_+^2}{l^2}\right), 
   \nonumber \\
&& \tilde S=\frac{r_+^n \mbox{Vol}(S^n)}{4G}.
\end{eqnarray}
The total entropy of the SdS solution (\ref{2eq1}) is the sum of the cosmological
horizon entropy $S$ and black hole horizon entropy $\tilde S$. However, as noticed
in Ref.~\cite{Cai1}, if one uses the BBM mass (\ref{2eq5}), the black hole horizon entropy
$\tilde S$ cannot be expressed by a form like the Cardy-Verlinde formula. Here we report
that if we adopt the mass definition due to Abbott and Deser~\cite{AD},  the entropy $\tilde S$
can also be rewritten in a Cardy-Verlinde form\footnote{We will make some remarks on the 
possible implications of this result in Sec.~5.}. The AD mass $\tilde E$ of the SdS solution 
is~\cite{AD} 
\begin{equation}
\label{2eq11}
\tilde E =M=\frac{r_+^{n-1}}{\omega_n}\left(1-\frac{r_+^2}{l^2}\right),
\end{equation}
which is expressed in terms of the black hole horizon radius $r_+$.
 In this case, the associated 
Casimir energy $\tilde E_c$, defined as $\tilde E_c =(n+1)\tilde E-n\tilde T \tilde S$,
is
\begin{equation}
\label{2eq12}
\tilde E_c =\frac{2n r_+^{n-1}\mbox{Vol}(S^n)}{16\pi G},
\end{equation}
and the extensive part of energy is 
\begin{equation}
\label{2eq13}
2\tilde E-\tilde E_c =-\frac{2nr_+^{n+1} \mbox{Vol}(S^n)}{16\pi G l^2}.
\end{equation}
It is then easy to see that the black hole horizon entropy in (\ref{2eq10}) can 
be rewritten as
\begin{equation}
\label{2eq14}
\tilde S= \frac{2\pi l}{n}\sqrt{\tilde E_c|2\tilde E-\tilde E_c|},
\end{equation}
once again, a form of the Cardy-Verlinde formula. However, here it is worthwhile to notice
that the extensive part (\ref{2eq13}) of energy is negative. 

The AD mass obeys the first law of thermodynamics associated with the black hole horizon
\begin{equation}
\label{2eq15}
d\tilde E=\tilde Td\tilde S.
\end{equation}
Considering the fact that $\tilde E =-E =M$ and combining (\ref{2eq15}) and (\ref{2eq9}),
one has
\begin{equation}
\label{2eq16}
\tilde Td\tilde S +TdS=0.
\end{equation}
However, it is not suitable to see it as the first law 
of thermodynamics for the SdS black hole solution, because in general both the two temperatures
associated with the cosmological and black hole horizons are not equal to each another, and
then the SdS black hole solution is unstable quantum-mechanically. Instead an appropriate
way is to rewrite (\ref{2eq16}) as
\begin{equation}
\label{2eq17}
\tilde \kappa d\tilde A+\kappa dA=0,
\end{equation} 
where $\tilde \kappa (\kappa)$ and $\tilde A (A)$ are the surface gravity and  area
of black hole (cosmological) horizon, respectively. Equation (\ref{2eq17}) is just the first
law of SdS black hole mechanics.  For another derivation of the first law of four-dimensional
SdS black holes see \cite{GH}.

\sect{Reissner-Nordstr\"om-de Sitter Black Holes }

In this section we extend the previous discussions to the case of Reissner-Nordstr\"om-dS
black hole solutions, whose metric is
\begin{eqnarray}
\label{3eq1}
&& ds^2 = -f(r) dt^2 +f(r)^{-1}dr^2 +r^2 d\Omega_n^2, \nonumber \\
&&~~~~~~ f(r)=1 -\frac{\omega_n M}{r^{n-1}} +\frac{n \omega_n^2 Q^2}{8(n-1) r^{2n-2}}
     -\frac{r^2}{l^2},
\end{eqnarray}
where $Q$ is the electric/magnetic charge of Maxwell field. For general $M$ and $Q$, the 
equation $f(r)=0$ may have four real roots. Three of them are real: the largest one is the 
cosmological horizon $r_c$,
the smallest is the inner (Cauchy) horizon of black hole, the middle one is the outer horizon
$r_+$ of black hole.  And the fourth is negative and has no physical meaning. The classification
of the RNdS solution has been made in Ref.~\cite{Roman} (see also \cite{CJS}). Some thermodynamic
quantities associated with the cosmological horizon are
\begin{eqnarray}
\label{3eq2}
&& T= \frac{1}{4\pi r_c} \left(-(n-1) +(n+1)\frac{r_c^2}{l^2}
    +\frac{n\omega_n^2 Q^2}{8 r_c^{2n-2}}\right), \nonumber \\
&& S =\frac{r_c^n\mbox{Vol}(S^n)}{4G}, \nonumber \\
&& \phi =-\frac{n}{4(n-1)}\frac{\omega_n Q}{r_c^{n-1}},
\end{eqnarray}
where $\phi$ is the chemical potential conjugate to the charge $Q$. In the BBM prescription,
the gravitational mass, subtracted the anomalous Casimir energy, of the RNdS solution is
\begin{equation}
\label{3eq3}
E=-M =-\frac{r_c^{n-1}}{\omega_n} \left (1 -\frac{r_c^2}{l^2} +
    \frac{n\omega_n^2 Q^2}{8(n-1)r_c^{2n-2}}\right).  
\end{equation}
The Casimir energy $E_c$, defined as $E_c =(n+1) E-nTS-n\phi Q$ in this case, is
found to be
\begin{equation}
\label{3eq4}
E_c=-\frac{2nr_c^{n-1}\mbox{Vol}(S^n)}{16\pi G},
\end{equation}
which has a same form as the case of SdS solution. Thus we can see that the entropy (\ref{3eq2})
of the cosmological horizon can be rewritten as\footnote{For related discussion, see 
also \cite{Med}, from another angle.}
\begin{equation}
\label{3eq5}
S=\frac{2\pi l}{n}\sqrt{|E_c|(2(E-E_q)-E_c)},
\end{equation}
where
\begin{equation}
E_q = \frac{1}{2}\phi Q =-\frac{n}{8(n-1)}\frac{\omega_n Q^2}{r_c^{n-1}}. 
\end{equation} 
We note that  the entropy expression (\ref{3eq5}) has a similar form as the case of
charged AdS black holes~\cite{Cai2}, there it is also found that the energy of electromagnetic
field has to be subtracted from the total energy. Furthermore, the first law of thermodynamics
of the cosmological horizon is
\begin{equation}
\label{3eq7}
dE =TdS +\phi dQ.
\end{equation}

For the black hole horizon, associated thermodynamic quantities are
\begin{eqnarray}
\label{3eq8}
&& \tilde T=\frac{1}{4\pi r_+}\left( (n-1) -(n+1)\frac{r_+^2}{l^2} -\frac{n\omega_n^2 Q^2}
   {8r_+^{2n-2}}\right), \nonumber \\
&& \tilde S=\frac{r_+^n \mbox{Vol}(S^n)}{4G}, \nonumber \\
&& \tilde \phi =\frac{n}{4(n-1)}\frac{\omega_n Q}{r_+^{n-1}}. 
\end{eqnarray}
The AD mass of RNdS solution can be expressed in terms of black hole horizon radius
$r_+$ and charge $Q$,
\begin{equation}
\label{3eq9}
\tilde E =M =\frac{r_+^{n-1}}{\omega_n} \left (1-\frac{r_+^2}{l^2} +
   \frac{n\omega_n^2 Q^2}{8(n-1)r_+^{2n-2}}\right).
\end{equation}
In this case, the Casimir energy, defined as $\tilde E_c =(n+1)\tilde E -n\tilde T\tilde
 S-n\tilde \phi Q$, is 
\begin{equation}
\label{3eq10}
\tilde E_c =\frac{2n r_+^{n-1}\mbox{Vol}(S^n)}{16\pi G},
\end{equation}
and the black hole entropy $\tilde S$ can be rewritten as
\begin{equation}
\label{3eq11}
\tilde S =\frac{2\pi l}{n}\sqrt{\tilde E_c |2(\tilde E-\tilde E_q)-\tilde E_c|},
\end{equation}
where 
\begin{equation}
\tilde E_q =\frac{1}{2}\tilde \phi Q=\frac{n\omega_n Q^2}{8(n-1)r_+^{n-1}},
\end{equation}
which is the energy of electromagnetic field outside the black hole horizon. Thus
we demonstrate that the black hole horizon entropy of RNdS solution can be expressed 
in a form as the Cardy-Verlinde formula. The AD mass and thermodynamic quantities of black hole
horizon satisfy the first law,
\begin{equation}
\label{3eq13}
d\tilde E =\tilde Td\tilde S +\tilde \phi dQ.
\end{equation}
Considering $\tilde E =-E =M$ and combining (\ref{3eq7}) with (\ref{3eq13}),
one has 
\begin{equation}
\label{3eq14}
\tilde Td\tilde S +TdS +\tilde \phi dQ +\phi dQ=0.
\end{equation}
As the case of SdS solution, we can rewrite the above as
\begin{equation}
\label{3eq15} 
\tilde \kappa d\tilde A +\kappa dA +8\pi G \triangle \phi dQ=0,
\end{equation}
where $\triangle\phi \equiv \tilde \phi +\phi=\frac{n\omega_nQ}{4(n-1)}
\left(\frac{1}{r_+^{n-1}}-\frac{1}{r_c^{n-1}} \right) $ is the chemical potential difference
between the black hole horizon and the cosmological horizon. Equation (\ref{3eq15}) is the
first law of RNdS black hole mechanics.


\sect{Kerr-de Sitter Black Holes}

The higher dimensional Kerr-AdS black hole solution has been given in Ref.~\cite{Hawk}, from 
which one can obtain an $(n+2)$-dimensional Kerr-dS black hole solution with a single angular
momentum parameter by replacing $l^2$ by $-l^2$. The metric of KdS solution is
\begin{eqnarray}
\label{4eq1}
&& ds^2=- \frac{\triangle_r}{\rho^2}\left(dt-\frac{a}{\Xi}\sin^2\theta d\phi\right)^2 
     +\frac{\rho^2}{\triangle_r}dr^2 +\frac{\rho^2}{\triangle_{\theta}}d\theta^2
      \nonumber \\
&&~~~~~~~~- \frac{\triangle_{\theta}\sin^2\theta}{\rho^2}\left(adt 
        -\frac{r^2+a^2}{\Xi}d\phi\right)^2 + r^2\cos^2\theta d\Omega_{n-2}^2,
\end{eqnarray}
where 
\begin{eqnarray}
\label{4eq2}
&& \triangle_r =(r^2+a^2)\left(1-\frac{r^2}{l^2}\right) -2M r^{3-n}, \ \
  \triangle_\theta =1+\frac{a^2}{l^2}\cos^2\theta, \nonumber \\
&& \Xi= 1+\frac{a^2}{l^2}, \ \ \rho^2 =r^2 +a^2\cos^2\theta.
\end{eqnarray}
The KdS solution has the same horizon structure as the RNdS solution. The cosmological horizon
has associated thermodynamic quantities
\begin{eqnarray}
\label{4eq3}
&& T =\frac{r_c}{4\pi (r_c^2 +a^2)}\left(-(n-1)\left(1-\frac{a^2}{l^2}\right)
     -(n-3)\frac{a^2}{r_c^2}+(n+1)\frac{r_c^2}{l^2}\right), 
      \nonumber \\
&& S= \frac{\mbox{Vol}(S^n)}{4G\Xi}r_c^{n-2}(r_c^2+a^2), \nonumber \\
&& J=\frac{a\mbox{Vol}(S^n)}{8\pi G\Xi^2}r_c^{n-3}(r_c^2+a^2)\left(1-\frac{r_c^2}{l^2}\right),
      \nonumber \\
&& \Omega =-\frac{a \Xi}{r_c^2+a^2},
\end{eqnarray}
where $J$ is the angular momentum  of the KdS solution in the BBM prescription~\cite{Dehg}
and $\Omega$ is the chemical potential conjugate to $J$, which can be explained as the
angular velocity of the cosmological horizon.  After subtracting the anomalous Casimir energy
of pure dS space in the rotating coordinates, the BBM mass of KdS 
is~\cite{Dehg}\footnote{In Ref.~\cite{Dehg} the gravitational mass and angular momentum 
are calculated, in the BBM prescription, for four-, five- and seven-dimensional KdS solutions,
respectively. We expect that the expression (\ref{4eq4}) and the angular momentum 
in (\ref{4eq3}) always hold in any dimension.}
\begin{equation}
\label{4eq4}
 E=- \frac{n\mbox{Vol}(S^n)}{16\pi G \Xi}r_c^{n-3}(r_c^2+a^2)
      \left(1-\frac{r_c^2}{l^2}\right),
\end{equation}
which is expressed in terms of the cosmological horizon radius $r_c$ and angular momentum 
parameter $a$. For the thermodynamics of cosmological horizon, we find that the
Casimir energy, defined as $E_c=(n+1)E-nTS -n\Omega J$, is
\begin{equation}
\label{4eq5} 
E_c=-\frac{2n \mbox{Vol}(S^n)}{16\pi G \Xi}r_c^{n-3}(r_c^2+a^2),
\end{equation}
and the extensive part of energy 
\begin{equation}
2E-E_c=\frac{2n r_c^{n-1}\mbox{Vol}(S^n)}{16\pi G\Xi l^2}(r_c^2+a^2).
\end{equation}
Thus it is easy to see that the entropy (\ref{4eq3}) of cosmological horizon can be re-expressed
as 
\begin{equation}
\label{4eq7}
S=\frac{2\pi l}{n}\sqrt{|E_c|(2E-E_c)},
\end{equation}
a completely same form as the case of SdS solution.  For the cosmological horizon, its first
law of thermodynamics is
\begin{equation}
\label{4eq8}
dE=TdS +\Omega dJ.
\end{equation}

 The AD mass $\tilde E$ and angular momentum $\tilde J$ and other thermodynamic quantities 
associated with the black hole horizon are
\begin{eqnarray}
&& \tilde T= \frac{r_+}{4\pi(r_+^2+a^2)}\left( (n-1)\left(1-\frac{a^2}{l^2}\right) 
            +(n-3)\frac{a^2}{r_+^2} -(n+1)\frac{r_+^2}{l^2}\right),
           \nonumber \\
&& \tilde S =\frac{\mbox{Vol}(S^n)}{4G\Xi}r_+^{n-2}(r_+^2+a^2), \nonumber\\
&& \tilde J = \frac{a\mbox{Vol}(S^n)}{8\pi G\Xi^2}r_+^{n-3}(r_+^2+a^2)
            \left(1-\frac{r_+^2}{l^2}\right), \nonumber \\
&& \tilde E = \frac{\mbox{Vol}(S^n)}{16\pi G\Xi}r_+^{n-3}(r_+^2+a^2)
            \left(1-\frac{r_+^2}{l^2}\right), \nonumber \\
&&\tilde\Omega = \frac{a\Xi}{r_+^2+a^2}.
\end{eqnarray}
Note that here we have  $\tilde E =-E$ and $\tilde J=J$. The AD mass and angular momentum
satisfy 
\begin{equation}
\label{4eq10}
d\tilde E = \tilde Td\tilde S + \tilde \Omega d\tilde J.
\end{equation}
Combining (\ref{4eq10}) with (\ref{4eq8}), we obtain the first law of KdS black hole mechanics,
\begin{equation}
\label{4eq11}
\tilde \kappa d\tilde A +\kappa dA +8\pi G \triangle \Omega dJ=0,
\end{equation}
where $\triangle \Omega \equiv \tilde \Omega +\Omega= a \Xi\left(\frac{1}{r_+^2+a^2}
 -\frac{1}{r_c^2+a^2}\right)$ is the angular velocity of black hole horizon with respect to
the cosmological horizon.  
 
For the thermodynamics of black hole horizon, the Casimir energy, defined as $\tilde E_c=
(n+1)\tilde E-n\tilde T\tilde S-n\tilde \Omega \tilde J$, is found to be
\begin{equation}
\tilde E_c= \frac{2n \mbox{Vol}(S^n)}{16\pi G \Xi}r_+^{n-3}(r_+^2+a^2),
\end{equation}
and the extensive part of energy is
\begin{equation}
2\tilde E-\tilde E_c=-\frac{2nr_+^{n-1}\mbox{Vol}(S^n)}{16\pi G\Xi l^2}(r_+^2+a^2).
\end{equation}
Once again, the extensive energy is negative. The entropy $\tilde S$ of black hole 
horizon can be rewritten as 
\begin{equation}
\label{4eq14}
\tilde S=\frac{2\pi l}{n}\sqrt{\tilde E_c|2\tilde E-\tilde E_c|},
\end{equation}
a form as the case of SdS black hole solutions. 

In the KdS solution (\ref{4eq1}), a special case is the one where $M=0$. In this case,
it can be shown that the metric (\ref{4eq1}) with $M=0$ describes a pure dS space in 
rotating coordinates~\cite{BF}. That is, the solution (\ref{4eq1}) with $M=0$ can be
changed into the pure dS space in the static coordinates (namely the solution (\ref{2eq1})
with $M=0$) by a coordinate transformation. For the pure dS space in rotating coordinates,
the cosmological horizon is still at $r_c=l$, the BBM mass (\ref{4eq4}) and angular momentum
in (\ref{4eq3}) vanish. The Hawking temperature and entropy are
\begin{eqnarray}
\label{4eq15}
&& T= \frac{1}{2\pi l}, \nonumber \\
&& S=\frac{l^n \mbox{Vol}(S^n)}{4G}.
\end{eqnarray}
The Casimir energy from (\ref{4eq5}) is 
\begin{equation}
E_c=- \frac{2n l^{n-1}\mbox{Vol}(S^n)}{16\pi G}. 
\end{equation}
These quantities are the same as those in the case of the static coordinates. Therefore the  entropy 
of cosmological horizon of the pure dS space in the rotating coordinates can also be given
using the Cardy-Verlinde formula.  Further, it seems to indicate  that  although the
vacuum states might be different in the static coordinates and rotating coordinates, the dual
CFT should give rise to the same thermo excitation for the pure dS space.

\sect{Conclusion and Discussion}

Holographic principle says that a theory with gravity in $D$ dimensions can be equivalent to
a theory in $(D-1)$ dimensions without gravity~\cite{Hooft}. The AdS/CFT 
correspondence~\cite{Mald} is a beautiful realization of the principle. Recently it has been
proposed that quantum gravity on a dS space is dual to a Euclidean CFT on a boundary of
the dS space (dS/CFT correspondence)~\cite{Stron1}. Unlike the AdS/CFT 
correspondence, however, the understanding to the dS/CFT correspondence so far acquired 
is quite incomplete.  In order to establish the dS/CFT correspondence, one has to first
collect more theoretic data. To well understand the gravity of asymptotically dS spaces is 
one of important topics.  

In a previous paper~\cite{Cai1}, adopting the BBM prescription~\cite{BBM} to compute 
boundary stress-energy tensor and conserved charges of asymptotically dS spaces, we have found
that the entropy associated with the cosmological horizon in the SdS solutions and topological 
dS solutions can be rewritten in terms of the Cardy-Verlinde formula, which is supposed
 to be an entropy formula of CFTs in any dimension. In this paper we have shown that
the conclusion also holds for the RNdS solutions and KdS solutions [see (\ref{3eq5}) and
(\ref{4eq7})].  This result therefore provides support of the dS/CFT correspondence.   

For spacetimes of black holes in dS spaces, however, the total entropy is the sum of black
hole horizon entropy and cosmological horizon entropy. If one uses the BBM mass of
the asymptotically dS spaces, the black hole horizon entropy cannot be expressed by a form
like the Cardy-Verlinde formula.  In this paper, we have found that if one uses the AD
prescription to calculate conserved charges of asymptotically dS spaces, the (SdS, RNdS, KdS)
black hole horizon entropy can also be rewritten [see (\ref{2eq14}), (\ref{3eq11}) and
(\ref{4eq14})] in a form of Cardy-Verlinde formula, which
indicates that the thermodynamics of black hole horizon in dS spaces can be also described by a
certain CFT.  
 
Our result seems to imply that we need two different CFTs associated respectively with the
black hole horizon and cosmological horizon to describe the dS black hole spacetimes. It looks
reasonable because generically the black hole horizon and cosmological horizon have different
temperatures, one do not expect that a same field theory can describe simultaneously the
thermodynamics of black hole horizon and of cosmological horizon. Our result is also 
reminiscent of the Carlip's claim~\cite{Carlip} that for black holes in any dimension the 
Bekenstein-Hawking entropy can be reproduced using the Cardy formula~\cite{Cardy}.  Carlip
obtained his result by considering general relativity on a manifold with boundary. He found
that the constraint algebra of general relativity may acquire a central extension, which can be 
calculated using covariant phase techniques. When the boundary is a (local) Killing horizon, a 
natural set of boundary conditions leads to a Virasoro subalgebra with a calculable central
charge. He then used conformal field theory methods to determine the density of states at the
boundary, which yields the expected entropy of black holes. The Carlip's method is also
applicable to the cosmological horizon. For the case of pure dS spaces, an analysis has been
made in \cite{LWu}. Applying the Carlip's method to black holes in de Sitter spaces,  obviously 
one  cannot give the total entropy using a same conformal field theory because in this method
the central charges associated with the black hole horizon and cosmological horizon are 
different, which also implies that one needs two different CFTs for the  black holes in
dS spaces.  As a result, the holography dual to the black holes in dS spaces might be
quite complicated. In order to establish the hologram for black holes in dS spaces, many
issues remain to be investigated.

\section*{Acknowledgments}
The author thanks Y.X. Chen, S. Hu, C.G. Huang, Q.G. Huang, M. Li, J.X. Lu,
 Y.S. Myung, C.J. Zhu and all participants of miniworkshop of superstring theory held 
in Hangzhou, China, Dec. 2001, for stimulating 
discussions of many related issues. This work was supported in part by a grant from Chinese 
Academy of Sciences.


\end{document}